# State of the Art Development on Solid-State Lithium Batteries

*L.J. Zhang*

**Lithium-ion batteries (LIBs) have aided wide range of applications, including the portable electronics, stationary storage and electric vehicles[1-8]. However, the safety risks associated with the use of organic electrolytes in lithium-ion batteries hinder its expansion of further application scenarios. Solid-state lithium batteries (SLBs) offer a promising avenue for the development of next-generation lithium-ion batteries with ultrahigh energy density and safety performance[9-15]. This review provides a quick overview of the state-of-the-art development of anode, cathode, solid electrolyte of SLBs and the observation of ion transport in the cell during the past half year in 2023. Other important developments for SLBs such as high safety and performance strategies have also been provided.**

## 1 Advances on cathode of SLBs

$TiS_2$-$TiO_2$ hybrid sheets cathode was developed to construct a novel solar Li ion battery with type II semiconductor heterostructure and lateral heterostructure geometry. Higher Li-ion insertion to $TiS_2$ was realized due to its higher Li binding energy (1.6 eV) compared to $TiO_2$ (1.03eV)[16]. The charging of lithium-ion full solid cell with light was interestingly demonstrated.

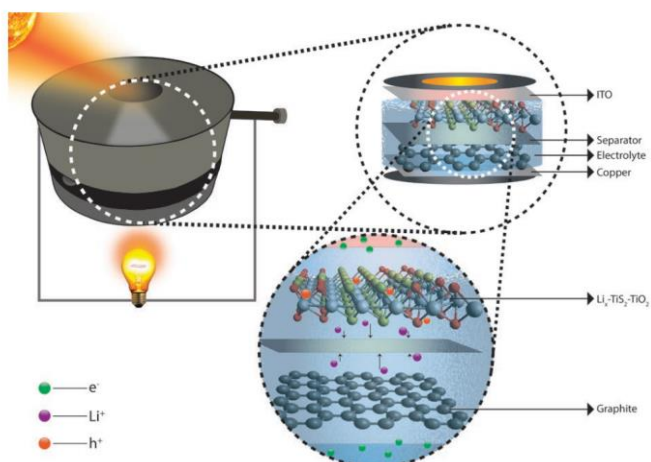

a photo-rechargeable lithium ion battery full cell using multifaceted layered cathode and graphite anode[16]. Copyright 2023 Wiley-VCH

Halides have rarely been used as cathode materials for lithium batteries (LBs) with liquid electrolytes. But for solid-state battery, halides can work as a promising active material for cathodes. $VX_3$ (X=Cl, Br, I) solid-state electrolytes have been recently reported to enable fast $Li^+$ insertion/extraction and high rate capability and stability of all-solid-state batteries[17]. $VCl_3$-$Li_3InCl_6$-C cathode, with $Li_3InCl_6$\$Li_6PS_5Cl$ solid state electrode(SSE) and Li anode, shows reversible charge capacity of 162.5 mAh g$^{−1}$ at 0.1C and 148 mAh g$^{−1}$ at 1C. For rate capability, the $VCl_3$-$Li_3InCl_6$-C cathode presents capacity retention about 84-85.7 % at 3C, 4C, and 6C (over 200 cycles). Results also shows that $VCl_3$-$Li_3InCl_6$-C cathode suffers a significant loss of capacity (capacity, 40 mAh g$^{−1}$ after 100 cycles) with sulfide $Li_6PS_5Cl$ SSE, indicating the compatibility between halide electrodes and electrolytes is a crucial factor in affecting the specific capacity of electrodes.

In addition, $LiCoO_2$-based SLBs with the guanidinium perchlorate ($GClO_4$) ferroelectrics as the cathode coatings are reported recently[18], which showed specific capacity of 210.6 mAh g$^{-1}$, reaching 91.6% performance of the liquid battery. This result is related to the single-domain state and upward self-polarization of $GClO_4$ coating.

## 2 Recent advances on anode of SLBs

Anode-free all-solid-state lithium battery is believed to be a promising strategy for extraordinary performance SLBs. Recent research[19] showed that an interconnected carbon paper compounded with a solid electrolyte achieved >5000 cycles life and an areal capacity of >8 mAh cm$^{-2}$.

Anodes based on metal oxides and non-toxic earth-abundant elements has showed significant application potential due to the lower costs. The nano-borate coated iron oxide anode was prepared recently for semi-solid-state bipolar LIBs, showing high energy density up to ~ 350 Wh kg$^{-1}$, outstanding power density (~6,700 W kg$^{-1}$), and prolonged ~2000 cycle life (75% capacity retention, 2 C) [20].

## 3   Developments on solid electrolyte of SLBs

### (1) polymer plus

Solid state electrolytes are being urgently studied due to their important role in improving battery energy density and safety[21-28]. Polymer electrolytes often suffer from low ionic conductivity at room temperature[28, 29]. Recently, graphene quantum dots (GQDs) modified poly (ethylene oxide) (PEO) electrolyte was developed and presented high ionic conductivity, excellent rate performance and cycling stability, which can be attributed to abundant hydroxyl groups and amino groups acting as Lewis base site to promote the dissociation of lithium salts and provide more ion pathways.

A single-ion polymer electrolyte, PAF-220-all-solid polymer electrolyte (PAF-220-ASPE)[30] , was developed with Bis(trifluoromethane)sulfonimide lithium (LiTFSI) infused PAF-220-Li and Poly(vinylidene fluoride-co-hexafluoropropylene)(PVDF-HFP) through solution casting and pressing-disc method[30]. The Li$^+$ conductivity of 0.501 mS cm$^{-1}$ can be attributed to reduced concentration polarization and lithium dendrite growth inhibition. Composite polymer electrolytes (CPEs) reached improved performance with 3% addition of active boron nitride nanosheets (BNNSs) [31], creating ionic conductivity of 6.7 × 10$^{-6}$ S cm$^{-1}$ at 30 °C, 2.77 × 10$^{-4}$ S cm$^{-1}$ at 50 °C. The LFP‖Li battery employing CPE/BNNS as electrolyte could present optimized rate performance (88.7 mA h g$^{-1}$ ,2.0C) and cycle stability (160.1 mA h g$^{-1}$ after 200 cycles, 0.5C 60 °C).

New research insights also involve the composition design of the solid electrolyte interphase (SEI). Li/electrolyte interface with enriched lithium fluoride was enabled by introducing copper polyphthalocyanine metal (CuPcLi) to PVDF-b-PTFE (PVT), which enhanced the Li-ion transport kinetics and created a lithium-ion conductivity of 0.8 mS cm$^{-1}$ and a lithium-ion transference number of 0.74 [32]. This PVT-10CuPcLi polymer electrolyte brings out improved performance, with lithium plating/stripping performance over 2000 h in a Li//Li half-cell and good capacity retention (92%) after 1000 cycles at 1 C at room temperature when paired with LiFePO$_4$.

Rubber-derived electrolytes with LiF-rich layer(ultraconformal layer) enabled rapid and stable lithium-ion transport across interfaces because the chemical connection between the electrolyte and lithium anode[33]. Stability over 300 cycles in a full cell and critical current density of 1.1 mA cm$^{-2}$ in lithium symmetric cells were observed.

Grafted MXenes based electrolyte, a solid polymer electrolyte, was also developed, where polyacrylonitrile grafted MXene (MXene-g-PAN) is employed as compatibilizer for poly(vinylidene fluoride-co-hexafluoropropylene) (PVHF)/PAN blends. With ionic conductivity of 2.17 × 10$^{-4}$ S cm$^{-1}$ and Li plating/stripping over 2500 h, the fabricated solid Li‖LiCoMnO$_4$ battery reaches a discharge voltage as high as 5.1 V [34].

A flexible composite electrolyte film (FSCEF) was also fabricated with ceramic and polyethylene oxide (PEO) polymer support, resulting in enhanced mechanical flexibility[35]. Polymer based quasi-solid-state electrolyte (QSE) also attracted attention these years, and high ionic conductivity of QSE (3.69 mS cm$^{-1}$) has been reported, which is related to significantly higher coordination strength of Li$^+$ on tertiary amine (−NR$_3$) group[36]. Ordered and fast ion transport resulting from this QSE enabled about 220 cycles (at ≈1.5 mA cm$^{-2}$ ) for Li‖NCM811 batteries.

MXene-SiO$_2$ nanosheets were also used to fabricate quasi-solid polymer electrolytes (QSPEs) with sandwich-structured polyacrylonitrile (PAN)[37], showing promising ionic conductivity of ~1.7 mS cm$^{-1}$ at 30 degrees C, and a low interfacial impedance. The capacity retention could reach 81.5% after 300 cycles (at 1.0 C, room temperature) for a Li‖LiFePO$_4$ quasi-solid-state lithium metal battery.

### (2) MOFs and small organic molecule system

Low ionic conductivity and high interfacial impedance are usually the main defects that cannot be ignored for Metal–organic frameworks (MOFs) based solid electrolytes. Recent researches have advanced this type electrolytes, and high performance (ion conductivity 1.04 × 10$^{-3}$ S cm$^{-1}$, Li$^+$-transference number 0.71) was realized via a novel hierarchical porous H-ZIF-8 solid-state electrolyte[38]. In addition, halloysite nanotubes (HNT) optimized H-ZIF-8/HNT electrolyte demonstrated further enhancement on electrochemical properties (ion conductivity $7.74 \times 10^{-3}$ S cm$^{-1}$, Li$^+$-transference number 0.84)[38]. It is worth mentioning that, equipped with H-ZIF-8/HNT, high capacity retention (84%, 104.16 mA h g$^{-1}$) after 200 cycles was also achieved by a Li/LiFePO$_4$ battery.

Biomimetic ion channels were constructed on quasi-solid-state electrolytes via UiO-66 with hollow structure[39], enabling incorporation of more lithium ions and superior ionic conductivity ($1.15 \times 10^{-3}$ S cm$^{-1}$) and lithium-ion transference number (0.70) at room temperature. This research suppressed Li dendrites through the creation of uniform ion flux.

1,3-dioxolane electrolytes with carbon nanotubes (CNTs) was used to fabricate SLBs, enabling excellent interfacial contact among CNTs, active materials and electrolytes and high lithium-ion diffusion efficiency, high energy density, as well as amazing rate performance[40].

**(3) Inorganic electrolytes**

Inorganic solid electrolytes (SE) such as halides are also promising candidates for SLIBs, although the lower ion conductivity compared to sulfides. A new lithium-metal-oxy-halide material, LiMOCl$_4$ (M=Nb, Ta), with high ionic conductivities of 10.4 ~12.4 mS cm$^{-1}$, was reported recently[41]. This electrolyte material also showed an outstanding rate capability (capacity retention 80 % at 5C/0.1C) when working as cathode SE, demonstrating the high practical application potential of oxyhalides.

Li$_2$S-P$_2$S$_5$-B$_2$S$_3$ electrolytes with a high critical current density of 1.65 mA cm$^{-2}$ and a long cycling life of over 300 h was developed[42]. Benefiting from "multi-layer mosaic like" interphase, suppression of Li dendrite growth was realized, and lifetime of SLIBs was ultimately prolonged.

Garnet-type Li$_7$La$_3$Zr$_2$O$_{12}$ (LLZ) materials usually suffer from the formation of insulating impurities related to high-temperature (above 1000 °C) sintering process, which is beneficial to its high Li-ion conductivity. It is noteworthy that low temperature treated LLZ materials have been developed, breaking this traditional limitation. Nanosized Li$_{6.5}$La$_3$Zr$_{1.5}$Ta$_{0.5}$O$_{12}$(LLZT) solid electrolytes have been successfully prepared and sintered at only 400~550°C[43], showing Li-ion conductivity of $1.03 \times 10^{-4}$ S cm$^{-1}$ and bulk-type areal discharge capacity of 0.831 mAh cm$^{-2}$. In addition, Li$_{6.5}$La$_{2.9}$Ca$_{0.1}$Zr$_{1.4}$Bi$_{0.6}$O$_{12}$ containing Li$_3$BO$_3$ has also been demonstrated as a promising cathode SE candidate[44], which can be sintered even at temperature as low as 750 °C with conductivity of $3 \times 10^{-4}$ S cm$^{-1}$ (25 °C).

A fluorinated quasi-solid-state electrolyte (QSSE) with lithium-rich layered oxide (LRLO) materials is prepared and presented an ionic conductivity of 6.4 x 10$^{-4}$ S cm$^{-1}$ (30 °C) and a 5.6 V electrochemical stable window[45]. In addition, LRLO/QSSE/Li batteries showed excellent rate performance with a high initial capacity for 209.7 mA h g$^{-1}$.

## 4 Lithium ion transport pathway

Understanding the lithium ion transport pathway is essential for all solid-state lithium-ion batteries, and in order to elucidate this process, a contrast enhanced neutron imaging method employing $^6$Li was reported[46]. Enhanced Li contrast was created via different Li configuration, high $^6$Li content in anode while natural lithium in the solid electrolyte, thus the visual observation of diffusion pathway of Li ions in a solid-state Li–S battery was enabled through operando neutron radiography and in situ neutron tomography.

Operando measurements were also used to reveal lithium dynamics of SLIBs[47]. Linear and nonlinear extreme-ultraviolet spectroscopies investigation for solid-state electrolyte surface lithium ions found blueshift relative to bulk absorption spectra, which is related to hybridized Li-s/Ti-d orbitals. First-principles calculation also show that low-frequency rattling modes suppression underlies the large interfacial resistance.

## 5 Towards Safe High-Performance Solid Batteries

Research about mechanism for SLBs plays an indispensable role in developing safe batteries with high performance. Lithium-ion fast transport plays a significant role in improving the performance of SLBs. Montmorillonite (OMMT) after organic modification process, which can be rich in the Lewis acid centers, was employed to enhance lithium salt dissociation[48]. With construction of lithium-ion 3D network channels, a new electrolyte LOPPM, LiTFSI/OMMT/PVDF/P(VDF-HFP)/PMMA, consists of polyvinylidene fluoride (PVDF),

poly(vinylidene fluoride-hexafluoro propylene) (P(VDF-HFP)) copolymer and polymerized methyl methacrylate (MMA), performing as an excellent polymer electrolyte with high ionic conductivity of $1.1 \times 10^{-3}$ S cm$^{-1}$. It is worth noting that a battery with the LOPPM electrolyte for secondary recycling reach an initial capacity of 123.9 mAh g$^{-1}$.

The insights of simulation computing have also recently developed. Large-scale molecular dynamics simulations was employed to reveal the atomistic pathways of lithium crystallization at the solid interfaces, which found a multi-step crystallization pathways mediated by interfacial lithium atoms[49].

Recently, a detailed structure study of the second phases, usually resulting from the sintering process of $Li_7La_3Zr_2O_{12}$ (LLZO) garnet solid electrolytes, was carried out with transmission electron microscopy and low-dose high-resolution imaging. Results showed that γ-LiAlO$_2$, α-Li$_5$AlO$_4$, and β-Li$_5$AlO$_4$ were three typical second phases, and a higher Li/Al ratio is more inclined to generate Li-rich second phases, which support more highly ionic conductive of LLZO[50].

Undoubtedly, the combination of multiple advanced characterization techniques and simulation calculations to gain a deeper understanding of electrolyte structure design and battery operation mechanism provides an essential and important avenue to promote the development of SLBs with higher performance and safety.